\documentstyle{article}
\input epsf

\begin{document}

\begin{center}
{\bf {MORPHOLOGY OF THE UNIVERSAL FUNCTION FOR THE CRITICAL CIRCLE MAP} }

{\bf {R.~Delbourgo and Brian G. Kenny} }

University of Tasmania, GPO Box 252-21, Hobart, AUSTRALIA 7001
\vspace{.2in}

\begin{abstract}
We describe the morphology of the universal function for the critical
(cubic) circle map at the golden mean, paying particular attention to the
birth of inflection points and their reproduction. In this way one can fully
understand its intricacies.
\end{abstract}

{\bf Introduction}
\end{center}

\noindent At the critical point, the circle map has a cubic order of
inflection. Near the golden mean winding number the self-reproducing
(renormalization group) properties of the Fibonacci mode-locked steps \cite%
{G} leads to the notion of a universal circle map function $g(x)$, which has
received a great deal of study in the literature \cite{S,FKS,ROSS}. It obeys
the compatible \cite{N} pair of functional equations, 
\begin{equation}
g\left( g(x)\right) =g(\alpha ^{2}x)/\alpha ,  \nonumber
\end{equation}%
\begin{equation}
g(g(x)/\alpha ^{2})=g(\alpha x)/\alpha ^{2}.
\end{equation}%
The universal circle map constant $\alpha =-1.288575..$ \cite{S} can be
determined to extremely high accuracy \cite{B} by examining the equations
over a restricted range of $x$. In this paper however we focus on the
behaviour of $g(x)$ over large $x$, in order to make sense of its
morphology. Some years ago we tried a similar analysis for the
period-doubling universal function \cite{DW}; our aim here is to comprehend
how the ever-increasing number of inflection points are spawned as we go out
to ever larger $|x|$. We shall see that they can be discovered directly by a
close look at the functional equations and their derivatives, starting from
previously known series of inflection points.

In the next section we quickly recapitulate the origin and character of the
functional equations and derive the principal features, including the first
family of inflection points. In the last section we locate the higher
families of inflection points of $g$ and explain how successive ones are
derived theoretically. In this way one can better appreciate the morphology
of the universal circle map function and comprehend its complicated
structure.

\begin{center}
{\bf {Origin and Main features} }
\end{center}

\noindent The approach to the irrational golden mean winding number in
nonlinear maps of an angular variable $\theta$ is normally made \cite{G} via
a succession of Fibonnaci number ratios: $F_{n-1}/F_n$ as $n\rightarrow\infty
$. In particular for the critical circle map, 
\begin{equation}
\theta \rightarrow\theta^{\prime}=f(\theta,\Omega)\equiv
\Omega+\theta-\sin(2\pi\theta)/2\pi,
\end{equation}
with a cubic inflection point at the origin, it is conventional to work out 
the superstable values of the driving frequencies $\Omega_n$ which
lead to the rational $F_{n-1}/F_n$, 
\begin{equation}
[f]^{F_{n-1}}(0,\Omega_n)-F_n=0.
\end{equation}
(Above we are using the notation, $[f]^N(x)$ to denote the $N$-fold composition, 
$\underbrace{f(f(\cdots f}_N(x)\cdots))$.) Correspondingly, 
\begin{equation}
\Delta\theta_n \equiv [f]^{F_{n-2}}(0,\Omega_n)-F_{n-1}
\end{equation}
is the nearest fixed point to the origin. The universal Feigenbaum constants 
$\delta,\alpha$ of the map are then obtained \cite{S} as

\begin{equation}
\delta = \lim_{n\rightarrow\infty}\Delta\Omega_n/\Delta\Omega_{n-1}=
-2.833612..;\qquad \Delta\Omega_n\equiv \Omega_n-\Omega_{n-1},
\end{equation}
\begin{equation}
\alpha = \lim_{n\rightarrow\infty}\Delta\theta_n/\Delta\theta_{n-1}
=-1.288575..
\end{equation}
Parenthetically, we note the numerical value $\Omega_\infty=0.6066610635..$

By considering the set of functions 
\begin{equation}
\phi _{r}(\theta )=\alpha ^{n}\left( [f]^{F_{n}}(\Omega _{n+r},\theta
/\alpha ^{n})-F_{n-1}\right) ,
\end{equation}%
in the limit of large $n$ and taking the limit as $r\rightarrow \infty $,
one may establish that the limiting (Shenker) function $\phi $, evaluated at 
$\Omega _{\infty }$, obeys {\em two} compatible functional equations, 
\begin{equation}
\phi \left( \phi (\theta )/\alpha \right) =\phi (\alpha \theta )/\alpha
^{2}\qquad {\rm and}
\end{equation}
\begin{equation}
\phi \left( \alpha \phi (\theta )\right) =\phi (\alpha ^{2}\theta )/\alpha .
\end{equation}
Numerical approximations to (8) indicate that the function $\phi(x)$ (called
$f(x)$ by Shenker)
is a monotonically increasing function of its argument; therefore the
alternative function $g(x)=\alpha \phi (x)$ decreases monotonically with $x$
and one readily establishes that it satisfies the universal equations (1,2).
At the same time one may easily show that the scaled function $G(x)=g(\rho
x)/\rho $ obeys the self-same equation pair (9,10); thus one is at liberty
to set the normalization scale at will. We shall fix the unique zero of $g(x)
$ at $x=1$, that is $g(1)=0$, and thereby find the value $g(0)\equiv \lambda
=1.2452..$ to be the intercept at the origin. Because of the cubic nature of
inflection, $g^{\prime }(0)=g^{\prime \prime }(0)=0$.

Plots of $g(x)$ in the ranges -8 to -4, -4 to +4, +4 to +8 are respectively
given in Figures 1, 2 and 3. These plots are approximate and they were
obtained from equation (8) by going to high order $(n=14)$ at the
accumulation point $\Omega _{\infty }$ and scaling appropriately in $x$, as
indicated above. (The range -8 to 8 was chosen so that we have
sufficiently many inflection points to make the investigation of the morphology
worthwhile, but not too many to confuse the subsequent discussion.)
Two significant points may be noted from these plots:

\begin{itemize}
\item they exhibit a series of inflection points which are self-similar and
which proliferate on ever smaller (and larger) scales as we go out in $x$,

\item the function behaves asymptotically as $g(x) \simeq \alpha x$.
\end{itemize}

The last fact is readily verified from the initial functional equations (1)
and (2); our goal is to understand the first fact, namely the origin and
families of inflection points. Before doing so let us derive a number of
useful facts about $g$ at particular locations, some of which are required
later. By evaluating (1) at $x=1/\alpha ^{2}$ and (2) at $x=1/\alpha $, one
readily discovers that 
\[
g(1/\alpha ^{2})=1,\qquad \qquad g(1/\alpha )=\alpha ^{2};
\]%
the former also follows from (2) evaluated at $x=1/\alpha ^{2}$. Taking (1)
and (2) at $x=1$, leads to 
\[
g(\alpha ^{2})=\alpha \lambda ,\qquad \qquad g(\alpha )=\alpha ^{2}\lambda ,
\]%
the former following from (1a) evaluated at $x=1/\alpha $. Finally, we note
that working out (1) and (2) at $x=0$, gives 
\[
g(\lambda )=\lambda /\alpha ,\qquad \qquad g(\lambda /\alpha ^{2})=\lambda
/\alpha ^{2}.
\]%
The latter condition demonstrates that $\lambda /\alpha ^{2}$ is the
(single) fixed point of the universal circle function! Further relations
between $g(\lambda \alpha ^{M})$ can be found by substituting suitable $%
x=\alpha ^{N}$ in eqs. (1), but we will not need them below.

As the next step, differentiate equations (1,2) to obtain the related pair, 
\begin{equation}
g^{\prime }(x)g^{\prime }\left( g(x)\right) =\alpha g^{\prime }(\alpha ^{2}x)
\end{equation}%
\begin{equation}
g^{\prime }(x)g^{\prime }\left( g(x)/\alpha ^{2}\right) =\alpha g^{\prime
}(\alpha x).
\end{equation}%
Remember that where $g$ has zero first derivative, its second derivative
also vanishes (because the map is cubic), so equations (11,12) provide
information about the inflection points. A number of interesting values of
the derivatives can be obtained immediately. Setting $x=1/\alpha ^{2}$ in
(11) and $x=1/\alpha $ in (12) yields 
\[
g^{\prime }(1/\alpha ^{2})=\alpha ,\qquad \qquad g^{\prime }(1/\alpha
)=\alpha 
\]%
and one can check the correctness of these values from the graphs. Also,
taking the limit as $x\rightarrow 0$ in (11,12), one deduces that 
\[
g^{\prime }(\lambda )=\alpha ^{5},\qquad \qquad g^{\prime }(\lambda /\alpha
^{2})=\alpha ^{3},
\]%
because the inflection point is cubic. By similar means one may establish
relations between derivatives of $g$ at various points $x=\alpha ^{M}\lambda 
$ which provide useful checks on numerical work.

But one can go further through the following observation: if $\xi $
corresponds to an inflection point, (11) and (12) ensure that $\alpha \xi $
and $\alpha ^{2}\xi $ are also inflection points; by induction one generates
an entire sequence of such points, viz. $\alpha ^{n}\xi ,\,\,n=0,1,2,\ldots $.
However equation (11) at $x=1$ and (12) at $x=0$, inform us that 
\[
g^{\prime }(\alpha ^{2})=0,\qquad \qquad g'(\alpha)=0,
\]
so we conclude that $\xi _{n}\equiv \alpha ^{n},\,\,n=1,2,3,\ldots $
represent a family of inflection points, seeded by $\xi _{1}=\alpha $; for
our purposes we shall regard them as the {\em primary} or {\em parent}
series. These points can be picked out in Table 1, where we have listed all
the inflection points that occur between $x=-8$ to +8 and they can also be
spotted in the figures as marked dots.  This table contains 
many more inflection points and the question is: how are they seeded? 
The answer will be provided in the next section.

\begin{center}
{\bf {Families of inflection points} }
\end{center}

\noindent Examination of the figures shows that not all inflection points
belong to the primary sequence $\alpha^N$. The first non-standard inflection
point occurs at $x\simeq 2.25$ and to see where this comes from we examine
first the seed values $x_N \equiv g^{-1}(\alpha^N)$, $N=1,2,\ldots$ . These
may be read off from the graph of $g$ and are tabulated in Table 2. At those
locations, 
\[
g^{\prime}\left(g(x_N)\right)=g^{\prime}(\alpha^N)=0,
\]
and from (11) we are assured that the set $\alpha^M x_N$, $M=2,3,\ldots$
will generate a family of inflection points, some of which may not be new.
[For example we note again that the location $x_2=1/\alpha$ simply generates
the primary or parent family.] The first daughter family, {\em which is new}%
, is given by the secondary series, $\alpha^2 x_1,\alpha^3 x_1,\ldots$;
another daughter family, which is also new, corresponds to the sequence, $%
\alpha x_3, \alpha^2 x_3,\ldots$, etc. and it starts off with $M=1$, by
virtue of equation (12). However the family $\alpha^n x_4$ is {\em not new}
because one may readily establish that $x_4 = \alpha x_1$ by inserting $x=x_1
$ in equation (12). New secondary families $\alpha^n x_5, \alpha^n x_6, \ldots$
etc. begin with $n=1$, again via equation (12).

While the parent and daughter sequences lead to two generations of
inflection points, some of which are listed in Table 1, this does not
exhaust the class of such points. Proceeding in a similar fashion to above,
define the granddaughter sequence, 
\begin{equation}
x_{MN}\equiv g^{-1}(\alpha^M x_N)=g^{-1}\left(\alpha^M g^{-1}(\alpha^N)
\right).
\end{equation}
Many of these points are not new; by manipulating (8a) and (8b)
appropriately, one may prove that some daughters are identical with or
related to their aunts. For instance, 
\[
\alpha x_{21}=x_3,\quad\alpha x_{31}=x_6,\quad\alpha x_{25}=x_7,  \quad
x_{15}=x_6,\quad x_{43}=\alpha x_6,\ldots 
\]
As well, some of these granddaughters are interrelated; for instance 
\[
\alpha x_{13}=x_{41},\quad\alpha x_{16}=x_{51}, \ldots
\]
Only the independent $\alpha^n x_{MN}$ occurring between -8 and +8 are
stated in Table 1.

Practically all the remaining missing points of inflection in that range
arise from the next generation, $x_{LMN}\equiv g^{-1}\left(\alpha^L x_{MN}
\right).$ Again we have only listed those that spawn new series. The final
`missing' point within -8 to 8, at $x\simeq -7.45$, is noteworthy because it
belongs to the `fourth' generation. By following through this procedure one
may track all the inflection points over any desired range, with each family
being generated from the previous one, though many are identified with
previous members; for example the point $\alpha^2 x_{241}$ is the same as
its aunt $\alpha x_{61}$. Within the restricted range -8 to 8, there are no
other inflection points beyond the ones noted in the table. An interesting
observation is that the number of inflection points between $\alpha^n$ and
$\alpha^{n+2}$ equals the Fibonnaci number $F_n$; while we can readily show 
that this holds for subcritical maps where $\alpha$ is the inverse of the 
golden mean ratio, an exact proof eludes us for critical cubic maps.

All told we can see how the functional equations reproduce their structure
in a self-similar way, becoming ever more intricate and extended at one and
the same time: for instance, the region between -7.7 and -7.4 looks very
much like the region between -2 and 2. This feature happens everywhere and
is not unlike the morphology of the period-doubling universal function. We
expect that the cubic nature of inflection is not vital for the validity of
this conclusion and that it is common to all scaling functional equations of
the type $[g]^N(x)\propto g(\rho x)$. The exquisite self-similarity on every
scale as we go out in $x$ means that it is effectively impossible to make an
analytical approximation to $g$ over a wide range of $x$; but of course over
a small, limited range it is always possible to do so and indeed this is a
sensible way to compute the scaling constant $\alpha$, by focussing on the
region $x=0$ to 1 say.

\begin{table}[h]
\caption{Positions of successive inflection points of $g(x)$ to two decimal
points, between $x=-8$ \& +8. ID is the interpretation of their origin in
the recursive notation $x_{MN\ldots}=g^{-1}(\protect\alpha^Mx_{N\ldots})$.}%
\vspace{.1in}
\par
\begin{tabular}{||r|r|r||r|r|r||r|r|r||}
\hline
$x$ & $g$ & ID & $x$ & $g$ & ID & $x$ & $g$ & ID \\ \hline
-7.99 & 10.60 & $\alpha^7 x_1$ & -4.80 & 6.43 & $\alpha^5x_1$ & 3.73 & -4.58
& $\alpha^4x_1$ \\ 
-7.76 & 10.10 & $\alpha^2x_{61}$ & -4.52 & 5.92 & $\alpha^3x_{13}$ & 4.12 & 
-5.12 & $\alpha x_6$ \\ 
-7.65 & 9.91 & $\alpha^2x_{161}$ & -4.39 & 5.71 & $\alpha x_{33}$ & 4.58 & 
-5.46 & $\alpha^6$ \\ 
-7.56 & 9.83 & $\alpha^5x_{13}$ & -4.27 & 5.62 & $\alpha^3x_3$ & 5.12 & -6.42
& $\alpha^2x_5$ \\ 
-7.45 & 9.63 & $\alpha x_{3313}$ & -3.98 & 5.26 & $\alpha x_5$ & 5.27 & -6.77
& $\alpha x_{143}$ \\ 
-7.40 & 9.56 & $\alpha^3x_{313}$ & -3.55 & 4.99 & $\alpha^5$ & 5.48 & -6.91
& $\alpha^4x_3$ \\ 
-7.35 & 9.52 & $\alpha^3x_{33}$ & -2.90 & 3.97 & $\alpha^3x_1$ & 5.66 & -7.22
& $\alpha^2x_{33}$ \\ 
-7.26 & 9.39 & $\alpha^2x_{233}$ & -2.58 & 3.42 & $\alpha x_3$ & 5.74 & -7.37
& $\alpha^2x_{133}$ \\ 
-7.09 & 9.30 & $\alpha^5x_3$ & -2.14 & 3.20 & $\alpha^3$ & 5.83 & -7.42 & $%
\alpha^4 x_{13}$ \\ 
-6.87 & 8.95 & $\alpha^2x_{43}$ & -1.29 & 2.08 & $\alpha$ & 5.99 & -7.64 & $%
\alpha x_{61}$ \\ 
-6.78 & 8.78 & $\alpha^2x_{143}$ & 0 & 1.25 & 1 & 6.19 & -7.78 & $\alpha^6
x_1$ \\ 
-6.62 & 8.69 & $\alpha^3 x_5$ & 1.66 & -1.62 & $\alpha^2$ & 6.52 & -8.29 & $%
\alpha^3x_{16}$ \\ 
-6.34 & 8.32 & $\alpha x_7$ & 2.25 & -2.67 & $\alpha^2x_1$ & 6.67 & -8.53 & $%
\alpha x_{36}$ \\ 
-5.90 & 8.06 & $\alpha^7$ & 2.76 & -3.08 & $\alpha^4$ & 6.85 & -8.65 & $%
\alpha^3x_6$ \\ 
-5.31 & 7.06 & $\alpha^2x_6$ & 3.30 & -4.07 & $\alpha^2x_3$ & 7.20 & -9.09 & 
$\alpha x_8$ \\ 
-5.07 & 6.63 & $\alpha^2x_{16}$ & 3.52 & -4.41 & $\alpha^2x_{13}$ & 7.60 & 
-9.40 & $\alpha^8$ \\ \hline
\end{tabular}%
\end{table}

\begin{table}[h]
\caption{The primary family of inflection points between -8 \& +8 and their
corresponding inverses (to two decimal points).}\vspace{.1in}
\par
\begin{tabular}{|r||r|r|r|r|r|r|r|r|}
\hline
$N$ & 1 & 2 & 3 & 4 & 5 & 6 & 7 & 8 \\ \hline
$\alpha^N$ & -1.29 & 1.66 & -2.14 & 2.76 & -3.55 & 4.58 & -5.90 & 7.60 \\ 
\hline
$x_N\equiv g^{-1}(\alpha^N)$ & 1.35 & -0.78 & 1.99 & -1.74 & 3.09 & -3.20 & 
4.92 & -5.58 \\ \hline
\end{tabular}%
\end{table}

%%%%%begin figure captions%%%%

\begin{center}
{\bf Figure Captions}
\end{center}

\noindent Figure 1. $g(x)$ from $x=-8$ to $x=-4$. This range contains  20
inflection points.

\noindent Figure 2. $g(x)$ from $x=-4$ to $x=+4$. This range contains  13
inflection points (only 8 points between -3 to 3).

\noindent Figure 3. $g(x)$ from $x=+4$ to $x=+8$. This range contains  15
inflection points.

%%%%%begin figures here%%%%%%%
\setlength{\unitlength}{1cm}

\newpage

\begin{figure}[tbp]
\begin{picture}(14,20)
\put (0.5,7.0){\epsfxsize=13cm \epsfbox{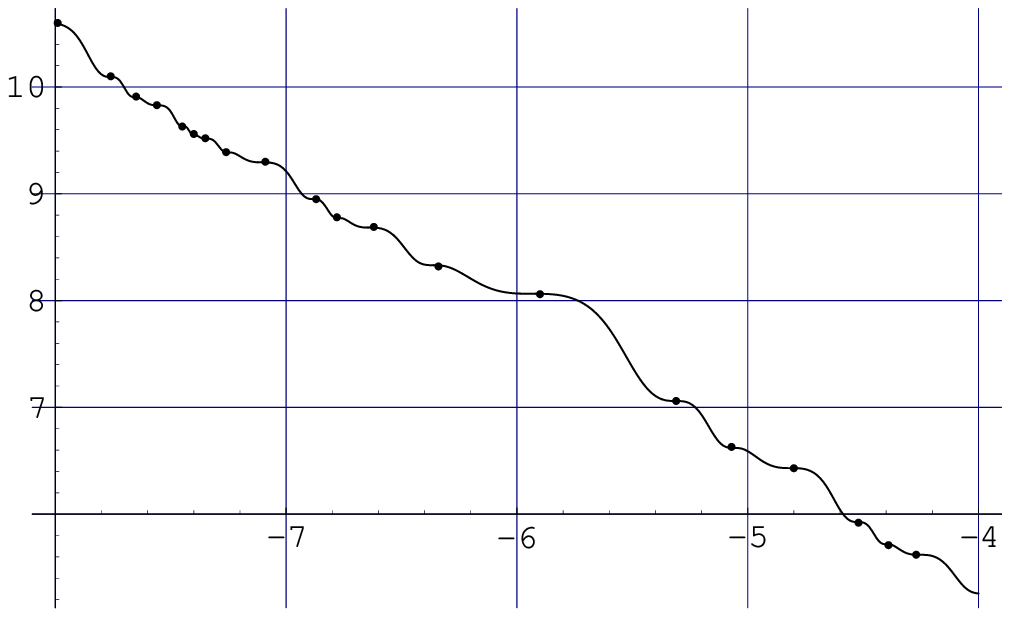}}
\end{picture}
\caption{R Delbourgo}
\label{fig:f1}
\end{figure}

\newpage

\begin{figure}[tbp]
\begin{picture}(14,20)
\put (0.5,7.0){\epsfxsize=13cm \epsfbox{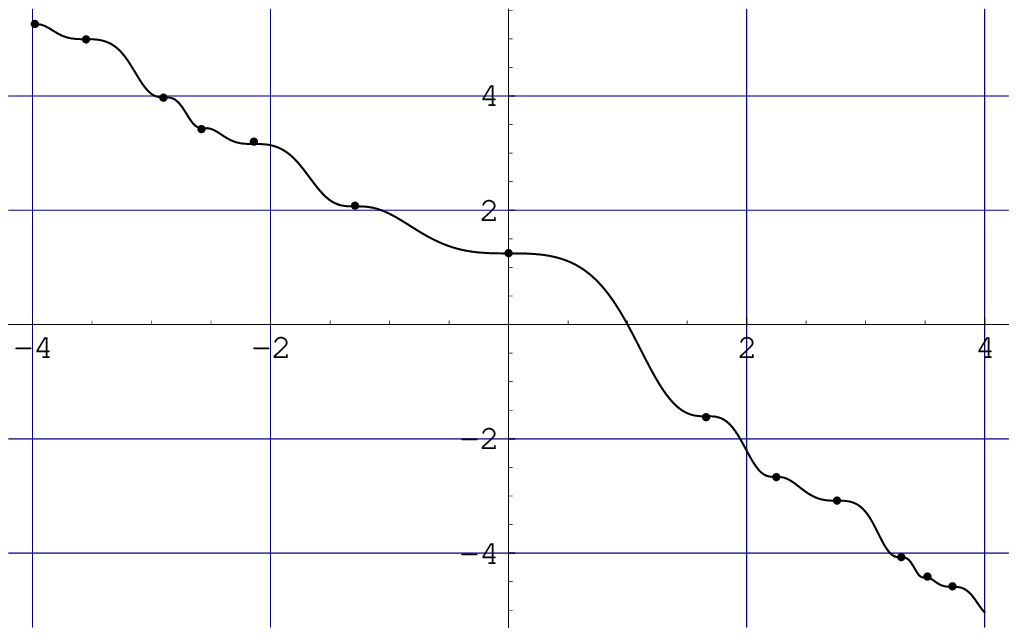}}
\end{picture}
\caption{R Delbourgo}
\label{fig:f2}
\end{figure}

\newpage

\begin{figure}[tbp]
\begin{picture}(14,20)
\put (0.5,7.0){\epsfxsize=13cm \epsfbox{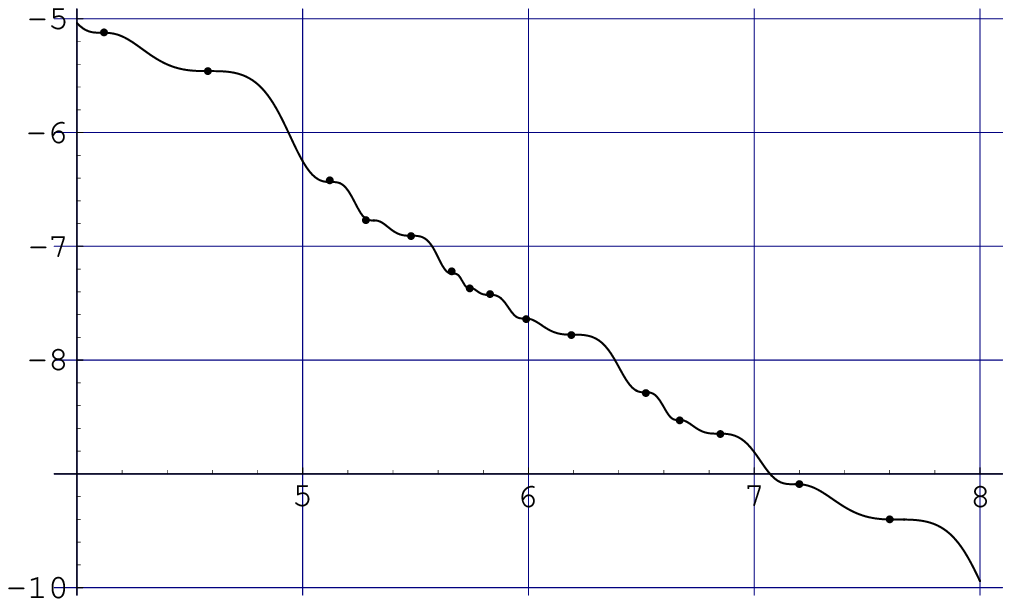}}
\end{picture}
\caption{R Delbourgo}
\label{fig:f3}
\end{figure}


\begin{thebibliography}{9}
\bibitem{N} M.~Nauenberg, On the fixed points for circle maps,
 {\em Phys. Lett.} {\bf 92A} (1982), 319-320 has
 shown the equivalence of the two equations, provided that $\phi(1/\alpha^2)= 
 1/\alpha^2$.
\end{thebibliography}
\end{document}